\documentclass[algorithms,article,submit,moreauthors,pdftex]{mdpi} 

\input{mysymbol.sty}
\usepackage{algorithm,algorithmic}
\usepackage{amsmath}%
\usepackage{MnSymbol}%
\usepackage{wasysym}%

\firstpage{1} 
\pubvolume{xx}
\issuenum{1}
\articlenumber{5}
\pubyear{2020}
\copyrightyear{2020}
\history{Received: date; Accepted: date; Published: date}

\Title{Online Topology Inference from Streaming Stationary Graph Signals with Partial Connectivity Information}

\Author{Rasoul Shafipour $^{1, 2, *}$ and Gonzalo Mateos $^{2, *}$}

\AuthorNames{Rasoul Shafipour and Gonzalo Mateos}

\address{%
$^{1}$ \quad Microsoft, Redmond, USA.\\
$^{2}$ \quad Dept. of Electrical and Computer Engineering, University of Rochester, USA.}

\corres{Correspondence: rasoul.shafipour@microsoft.com; gmateosb@ece.rochester.edu}

\abstract{We develop online graph learning algorithms from streaming network data. Our goal is to track the (possibly) time-varying network topology, and effect memory and computational savings by processing the data on-the-fly as they are acquired. The setup entails observations modeled as stationary graph signals generated by local diffusion dynamics on the unknown network. Moreover, we may have a priori information on the presence or absence of a few edges as in the link prediction problem. The stationarity assumption implies that the observations' covariance matrix and the so-called graph shift operator (GSO -- a matrix encoding the graph topology) commute under mild requirements. This motivates formulating the topology inference task as an inverse problem, whereby one searches for a sparse GSO that is structurally admissible and approximately commutes with the observations' empirical covariance matrix. For streaming data said covariance can be updated recursively, and we show online proximal gradient iterations can be brought to bear to efficiently track the time-varying solution of the inverse problem with quantifiable guarantees. Specifically, we derive conditions under which the GSO recovery cost is strongly convex and use this property to prove that the online algorithm converges to within a neighborhood of the optimal time-varying batch solution. Numerical tests illustrate the effectiveness of the proposed graph learning approach in adapting to streaming information and tracking changes in the sought dynamic network.}


\begin{document}


\section{Introduction}\label{S:Introduction}
Network data supported on the vertices of a graph $\ccalG$ representing pairwise interactions among entities are nowadays ubiquitous across disciplines spanning engineering as well as social and the bio-behavioral sciences; see e.g.,~\cite[Ch. 1]{kolaczyk2009book}. Such data can be conceptualized as graph signals, namely high-dimensional vectors with correlated entries indexed by the nodes of $\ccalG$. Graph-supported signals abound in real-world applications of complex systems, including vehicle congestion levels over road networks, neurological activity signals supported on brain connectivity networks, COVID-19 incidence in different geographical regions linked via mobility or social graphs, and fake news that spread on online social networks. Efficiently mining information from unprecedented volumes of network data promises to prevent or limit the spread of epidemics and diseases, identifying trends in financial markets, learning the dynamics of emergent social-computational systems, and also protect critical infrastructure including the smart grid and the Internet’s backbone network~\cite{kostas2014spmag}. In this context, the goal of graph signal processing (GSP) is to develop information processing algorithms that fruitfully exploit the relational structure of said network data~\cite{Ortega2018ProcIEEE}. However, oftentimes $\ccalG$ is not readily available and a first key step is to use nodal observations (i.e., measurements of graph signals) to identify the underlying network structure, or, a useful graph model that facilitates signal representations and downstream learning tasks; see~\cite{topoid_tutorial,dong2019learning} for recent tutorials on graph learning with a signal processing flavor and~\cite[Ch. 7]{kolaczyk2009book} for a statistical treatment. Recognizing that many of these networks are not only unknown but also \emph{dynamic}, there is a growing need to develop \emph{online} graph learning algorithms that can process network information streams in an efficient fashion.

\subsection{Identifying the structure of network diffusion processes}\label{Ss:Prob_statement}

Consider a weighted, undirected graph $\ccalG$ consisting of a node set $\ccalN$ of cardinality $N$, and symmetric adjacency matrix $\bbA\in\reals_+^{N\times N}$ with entry $A_{ij}\!=\!A_{ji}\!\neq\!0$ denoting the edge weight between node $i$ and node $j$. We assume that $\ccalG$ contains no self-loops; i.e., $A_{ii}\!=\!0$. The adjacency matrix encodes the topology of $\ccalG$. One could generically define a \emph{graph-shift operator} (GSO) $\bbS\in\reals^{N\times N}$ as any matrix capturing the same sparsity pattern as $\bbA$ on its off-diagonal entries. Beyond $\bbA$, common choices for $\bbS$ are the combinatorial Laplacian $\bbL:=\diag(\bbA\bbone)-\bbA$ as well as their normalized counterparts~\cite{Ortega2018ProcIEEE}. Henceforth we focus on $\bbS\!=\!\bbA$ and aim to recover the \emph{sparse} adjacency matrix of the unknown graph $\ccalG$. Other GSOs can be accommodated in a similar fashion. The graph can be dynamic with a slowly time-varying GSO $\bbS_t$, $t=1,2,\ldots$ (see also Section \ref{S:Online}), but for now we omit any form of temporal dependency to simplify exposition.

Our goal is to develop an online framework to estimate sparse graphs that explain the structure of a class of streaming random signals. At some time instant, let $\bby=[y_1,\ldots,y_N]^T \in\mbR^N$ be a zero-mean graph signal in which the $i$th element $y_i$ denotes the signal value at node $i$ of the \emph{unknown graph} $\ccalG$ with shift operator $\bbS$. Further consider a zero-mean white signal $\bbx$. We state that the graph $\bbS$ represents the structure of the signal $\bby\in\mbR^N$ if there exists a diffusion process in the GSO $\bbS$ that produces the signal $\bby$ from the input signal $\bbx$~\cite{segarra2016topoidTSP16}, that is

\begin{equation}\label{eqn_diffusion}
	\bby  = \textstyle  \alpha_0 \prod_{l=1}^{\infty} (\bbI_N-\alpha_l \bbS) \bbx
	= \sum_{l=0}^{\infty}\beta_l \bbS^l\,\bbx.
\end{equation}
Under the assumption that $\bbC_\bbx=\E{\bbx\bbx^T}=\bbI_N$ (identity matrix), \eqref{eqn_diffusion} is equivalent to the \emph{stationarity} of $\bby$ in $\bbS$; see e.g., \cite[Def. 1]{marques2016stationaryTSP16},~\cite{perraudinstationary2016},~\cite{girault_stationarity}. The product and sum representations in \eqref{eqn_diffusion} are common (and equivalent) models for the generation of random network processes. Indeed, any process that can be understood as the linear propagation of a white input through $\ccalG$ can be written in the form in \eqref{eqn_diffusion}, and subsumes heat diffusion, consensus and the classic DeGroot model of opinion dynamics as special cases. The justification to say that $\bbS$ represents the structure of $\bby$ is that we can think of the edges of $\ccalG$, i.e., those few non-zero entries in $\bbS$, as direct (one-hop) relations between the elements of the signal. The diffusion in \eqref{eqn_diffusion} modifies the original correlation by inducing indirect (multi-hop) relations in $\bbC_\bby=\E{\bby\bby^T}$. At a high level, the problem studied in this paper is the following~\cite{segarra2016topoidTSP16,pasdeloup2016inferenceTSIPN16}.\vspace{0.1cm} 

\noindent\textbf{Problem.} \emph{Recover the direct relations described by a sparse GSO $\bbS$ from a set $\{\bby_t\}_{t=1}^T$
of $T$ independent samples of the random signal $\bby$ adhering to \eqref{eqn_diffusion}.}

We consider the challenging setup when we have no knowledge of the diffusion coefficients $\{\alpha_l\}$ (or $\{\beta_l\}$), nor do we get to observe the specific realizations of the driving inputs $\{\bbx_t\}_{t=1}^T$. The stated inverse problem is severely underdetermined and nonconvex. It is underdetermined because for every observation $\bby$ we have the same number of unknowns in the input $\bbx$ on top of the unknown diffusion coefficients and the shift $\bbS$, the latter being the quantity of interest. The problem is nonconvex because the observations depend on the product of our unknowns and, notably, on powers of $\bbS$ [cf. \eqref{eqn_diffusion}]. Our main contribution is to develop novel algorithms to address these technical challenges in several previously unexplored settings.

\subsection{Technical approach and paper outline}\label{Ss:Paper_outline}

We tackle the network topology inference problem of the previous section in two fundamental settings. We start in Section \ref{S:prelim_problem} with the batch (offline) case where observations $\{\bby_t\}_{t=1}^T$ are all available for joint processing as in~\cite{segarra2016topoidTSP16,pasdeloup2016inferenceTSIPN16}. Here though we exploit the implications of stationarity from a different angle which leads to a formulation amenable to efficient solutions via proximal gradient (PG) algorithms; see e.g.,~\cite{Boyd2014Proximal,beck}. Specifically, as we show in Section \ref{Ss:Stationarity_Revisited} stationarity implies that the covariance matrix $\bbC_\bby$ of the observations commutes with $\bbS$ under mild requirements; see also~\cite{marques2016stationaryTSP16}. This motivates formulating the topology inference task as an inverse problem, whereby one searches for a sparse $\bbS$ that is structurally admissible and approximately commutes with the observations' empirical covariance matrix $\hbC_\bby$. In~\cite{segarra2016topoidTSP16,pasdeloup2016inferenceTSIPN16}, the algorithmic issues were not thoroughly studied and relied mostly on off-the-shelf convex solvers whose scalability could be challenged. Unlike~\cite{DSWonlinetopoid} but similar to link prediction problems~\cite[Ch. 7.2]{kolaczyk2009book},~\cite{RG_topoID_camsap_online}, here we rely on a priori knowledge about the presence (or absence) of a few edges; conceivably leading to simpler algorithmic updates and better recovery performance. We may learn about edge status via limited questionnaires and experiments, or, we could perform edge screening prior to topology inference~\cite{screening_camsap17}. The batch PG algorithm developed in Section \ref{Ss:Batch_Proximal} to recover the network  topology is computationally more efficient than existing methods for this (and related) problem(s). It also serves as a fundamental building block to construct online iterations, which constitutes the second main contribution of this paper.

Focus in Section \ref{S:Online} shifts to online recovery of $\bbS$ from \emph{streaming} signals $\{\bby_1,\ldots, \bby_t, \bby_{t+1},\ldots \}$, each of them adhering to the generative model in \eqref{eqn_diffusion}. For streaming data the empirical covariance estimate $\hbC_{\bby,t}$ can be updated recursively, and in Section \ref{Ss:algo_construction} we show online PG iterations can be brought to bear to efficiently track the time-varying solution of the inverse problem with quantifiable (non-asymptotic) guarantees. Different from~\cite{DSWonlinetopoid}, the updates are simple and devoid of multiple inner loops to compute expensive projections. Moreover, we establish convergence to within a neighborhood of the optimal time-varying batch solution as well as dynamic regret bounds (Section \ref{Ss:Convergence}) by adapting results in~\cite{Liam_online}. The algorithm and analyses of this paper are valid even for dynamic networks, i.e., if the GSO $\bbS_t$ in \eqref{eqn_diffusion} is (slowly) time-varying. Indeed, we examine how the variability and eigenvectors of the underlying graph as well as the diffusion filters' frequency response influence the size of the convergence radius (or misadjustment in the adaptive filtering parlance). Numerical tests in Section~\ref{S:numerical_results}  corroborate the efficiency and effectiveness of the proposed algorithm in adapting to streaming information and tracking changes in the sought dynamic network. Concluding remarks are given in Section \ref{S:conclusions}.

\subsection{Contributions in context of prior related work}

Early topology inference approaches in the statistics literature can be traced back to the problem of (undirected) graphical model selection~\cite[Ch. 7]{kolaczyk2009book},~\cite{topoid_tutorial,dong2019learning}. Under Gaussianity assumptions, this line of work has well-documented connections with covariance selection~\cite{Dempster1973} and sparse precision matrix estimation~\cite{GLasso2008,Lake10discoveringstructure,egilmez2017jstsp}, as well as neighborhood-based sparse linear regression~\cite{meinshausen06}. Recent GSP-based network inference frameworks postulate that the network exists as a latent underlying structure, and that observations are generated as a result of a network process defined in such a graph~\cite{DongLaplacianLearning,MeiGraphStructure, Kalofolias2016inference_smoothAISTATS16,segarra2016topoidTSP16,pasdeloup2016inferenceTSIPN16,thanou17}.

Different from~\cite{DongLaplacianLearning,Kalofolias2016inference_smoothAISTATS16,sandeep_icassp17,mike_icassp17} that infer structure from signals assumed to be smooth over the sought undirected graph, here
the measurements are assumed related to the graph via filtering [cf. \eqref{eqn_diffusion} and the opening discussion in Section \ref{S:prelim_problem}]. Few works have recently built on this rationale to identify a symmetric GSO given its eigenvectors, either assuming that the input is white~\cite{segarra2016topoidTSP16,pasdeloup2016inferenceTSIPN16} -- equivalently implying $\bby$ is graph stationary~\cite{girault_stationarity,marques2016stationaryTSP16,perraudinstationary2016}; or, colored~\cite{RSSSAMGM_TSP18,RasSantAntGonch2017icassp_SymmetricTopoInfer}.
Unlike prior \emph{online} algorithms developed based on the aforementioned graph spectral domain design~\cite{DSWonlinetopoid,RG_topoID_camsap_online}, here we estimate the (possibly) time-varying GSO directly (without tracking its eigenvectors) and derive quantifiable recovery guarantees; see Remark \ref{R:comparisons_conference}. Recent algorithms for identifying topologies of time-varying graphs~\cite{cardoso2020financial,kalofolias2017icassp} operate in batch mode, they are non-recursive and hence their computational complexity grows linearly with time. While we assume that the graph signals are stationary, the online graph learning scheme in~\cite{vlaski2018online} uses observations from a  Laplacian-based, continuous-time graph process. Unlike~\cite{shen2017tensors} that relies on a single-pole graph filter~\cite{elvin_arma_gf}, the filter structure underlying \eqref{eqn_diffusion} can be arbitrary, but the focus here is on learning undirected graphs. Online PG methods were adopted for directed graph inference under dynamic structural equation models~\cite{BainganaInfoNetworks}, but lacking a formal performance analysis. The interested reader is referred to~\cite{giannakis2018pieee} for a comprehensive survey about topology identification of dynamic graphs with directional links. The recovery guarantees in Section \ref{Ss:Convergence} are adapted from the results in~\cite{Liam_online}, obtained therein in the context of online sparse subspace clustering. Convergence and dynamic regret analysis techniques have been recently developed to study solutions of time-varying convex optimization problems; see~\cite{emiliano2020streams} for a timely survey of this body of work. The impact of these optimization advances to dynamic network topology identification is yet to fully materialize, and this paper offers a first exploration in this direction.  

In closing, we note that information processing from streams of network data has been considered in recent work that falls under the umbrella of \emph{adaptive} GSP~\cite{dilorenzo2017adaptivegsp,dilorenzo2018adaptivegsp}. Most existing results pertain to online reconstruction of partially-observed streaming graph signals, which are assumed to be bandlimited with respect to a \emph{known} graph. An exception is~\cite{vassilis2019tsp}, which puts forth an online algorithm for joint inference of the dynamic network topology and processes from partial nodal observations.

\subsection{Notational conventions}

The entries of a matrix $\mathbf{X}$ and a (column) vector $\mathbf{x}$ are denoted by $X_{ij}$ and $x_i$, respectively. Sets are represented by calligraphic capital letters and $\reals_+$ denotes the non-negative real numbers. 
The notation $^T$ stands for matrix or vector transposition; $\mathbf{0}$ and $\mathbf{1}$ refer to the all-zero and all-one vectors; while $\bbI_N$ denotes the $N\times N$ identity matrix. For a vector $\bbx$, $\diag(\mathbf{x})$ is a diagonal matrix whose $i$th diagonal entry is $x_i$. 
The operators $\otimes$, $\odot$, $\circ$, and $\text{vec}(\cdot)$ stand for Kronecker product, Khatri-Rao (columnwise Kronecker) product, Hadamard (entrywise) product and matrix vectorization, respectively. The spectral radius of matrix $\bbX$ is denoted by $\lambda_{\max}(\bbX)$ and $\| \bbX \|_p$ stands for the $\ell_p$ norm of $\text{vec}(\bbX)$. Lastly, for a function $f$, denote $\|f\|_{\infty}\!=\!\sup _{\bbX}|f(\bbX)|$.
 
\section{Identifying graph topologies from observations of stationary graph signals} \label{S:prelim_problem}

We start with \emph{batch} (offline) estimation of a time-invariant network topology as considered in~\cite{segarra2016topoidTSP16}. This context is useful to better appreciate the alternative formulation in Section \ref{Ss:Stationarity_Revisited} and the proposed algorithm in Section \ref{Ss:Batch_Proximal}. Together, these two elements will be instrumental in transitioning to the \emph{online} setting studied in Section \ref{S:Online}.  

To state the problem, recall the symmetric GSO $\bbS\in\reals^{N\times N}$ associated with the undirected graph $\ccalG$. Upon defining the vector of coefficients $\bbh:=[h_0,\ldots,h_{L-1}]^T\in\reals^L$ and the polynomial graph filter $\bbH:=\sum_{l=0}^{L-1} h_l \bbS^l\in \reals^{N\times N}$~\cite{Ortega2018ProcIEEE,elvin_arma_gf}, the Cayley-Hamilton theorem asserts that the model in \eqref{eqn_diffusion} can be equivalently reparameterized as

\begin{equation}\label{E:Filter_input_output_time}
	\bby  = \textstyle \big(\sum_{l=0}^{L-1}h_l \bbS^l\big)\,\bbx= \bbH \bbx, 
\end{equation}
for some particular $\bbh$ and filter order $L\leq N$. Since $\bbS$ is a local (one-hop) diffusion operator, $L$ specifies the (multi-hop) range of vertex interactions when generating $\bby$ from $\bbx$. It should now be clear that we are constraining ourselves to observations of signals generated via graph filtering. As we explain next, this assumption leads to a tight coupling between $\bbS$ and the second-order statistics of $\bby$.

The covariance matrix of $\bby=\bbH\bbx$ is given by [recall \eqref{E:Filter_input_output_time} and $\bbC_\bbx = \bbI_N$]

\begin{equation}\label{E:covariance_y_adaptive}
	\bbC_\bby:=\E{\bby\bby^T}=\E{\bbH\bbx(\bbH\bbx)^T}= \bbH \E{\bbx\bbx^T}\bbH	= \bbH^2.
\end{equation}
We relied on the symmetry of $\bbH$ to obtain the third equality, as $\bbH$ is a polynomial in the symmetric GSO $\bbS$. Using the spectral decomposition of $\bbS =\bbV\bbLam\bbV^T$ to express the filter as $\bbH = \sum_{l=0}^{L-1} h_l (\bbV\bbLam\bbV^T)^l =\bbV(\sum_{l=0}^{L-1} h_l \bbLam^l)\bbV^T$, we can readily diagonalize the covariance matrix in \eqref{E:covariance_y_adaptive} as 

\begin{equation}\label{eqn_diagonalize_covariance_adaptive}
	\bbC_\bby=\textstyle  \bbV\left(\sum_{l=0}^{L-1}h_l\bbLam^l\right)^2\bbV^T.
\end{equation}
Such a covariance expression is the requirement for a graph signal to be stationary in $\bbS$~\cite[Def. 2.b]{marques2016stationaryTSP16}. In other words, if $\bby$ is graph stationary (equivalently if $\bbC_\bbx = \bbI_N$), \eqref{eqn_diagonalize_covariance_adaptive} shows that the \textit{eigenvectors} of the shift $\bbS$, the  filter $\bbH$, and the covariance $\bbC_\bby$ are \textit{all the same}.  Thus given a batch of observations $\{\bby_t\}_{t=1}^T$ comprising independent realizations of \eqref{eqn_diagonalize_covariance_adaptive}, the approach in~\cite{segarra2016topoidTSP16} advocates: (i) forming the \textit{sample covariance} $\hbC_{\bby}=\frac{1}{T}\sum_{t=1}^{T}\bby_t\bby_t^T$ 
and extracting its eigenvectors $\hbV$ as spectral templates of $\ccalG$; then 
(ii) recover $\bbS$ that is optimal in some sense by estimating its eigenvalues $\bbLambda=\textrm{diag}(\lambda_1,\ldots,\lambda_N)$. Namely, one solves the inverse problem

\begin{equation}\label{topouE:general_problem}
	\min_{\bbLambda, \bbS \in \ccalS} \
	f ( \bbS) ,   \quad
	\text{subject to }\:d(\bbS,\hbV\bbLambda\hbV^T)\leq \epsilon
\end{equation}
which is convex for appropriate choices of the function $f(\bbS)$, the constraint set $\ccalS$ and the matrix distance $d(\cdot, \cdot): \reals^{N\times N}\times \reals^{N\times N} \mapsto \reals_{+}$. The tuning parameter $\epsilon>0$ accounts for the (finite) sample size-induced errors in estimating $\bbV$, and could be set to zero if the eigenvectors were perfectly known. The formulation \eqref{topouE:general_problem} entails a general class of network topology inference problems parameterized by the choices of $f$, $d$ and $\ccalS$; see~\cite{segarra2016topoidTSP16} for additional details on various application-dependent instances. Particular cases of \eqref{topouE:general_problem} with $\epsilon=0$ were independently studied in~\cite{pasdeloup2016inferenceTSIPN16}. 

In this paper we consider a different formulation from \eqref{topouE:general_problem}, which is amenable to efficient algorithms. For concreteness, we also make specific choices of $f$, $d$ and $\ccalS$ as described next.


\subsection{Revisiting stationarity for graph learning}\label{Ss:Stationarity_Revisited}

As a different take on the shared eigenspace property, observe that stationarity of $\bby$ also implies $\bbC_{\bby} \bbS = \bbS \bbC_{\bby}$, thus forcing the covariance $\bbC_{\bby}$ to be a polynomial in $\bbS$ [cf. \eqref{E:covariance_y_adaptive}]. This commutation identity holds under the pragmatic assumption that all the eigenvalues of $\bbS$ are simple and $\sum_{l=0}^{L-1}h_l\lam_i^l \neq 0$, for $i=1,\ldots,N$. It is also a natural characterization of (second-order) graph stationarity, requiring that second-order statistical information is shift invariant~\cite{marques2016stationaryTSP16}.  Since one can only estimate $\hbC_\bby$ from data $\{\bby_t\}_{t=1}^T$, our idea to recover a \emph{sparse} GSO is to solve [cf. \eqref{topouE:general_problem}]

\begin{equation}\label{topouE:general_problem_directed}
	\bbS^* := \argmin_{\bbS \in \ccalS} \
	\| \bbS\|_1 ,   \quad
	\text{subject to }\:\|\bbS\hbC_{\bby}-\hbC_{\bby} \bbS\|_F^2\leq \epsilon.
\end{equation}
The inverse problem \eqref{topouE:general_problem_directed} is intuitive. The constraints are such that one searches for an admissible $\bbS\in\ccalS$ that approximately commutes with the observations' empirical covariance matrix $\hbC_\bby$. To recover a sparse graph we minimize the $\ell_1$-norm of $\bbS$, the workhorse convex proxy to the intractable cardinality function $\|\bbS\|_0$ counting the number of non-zero entries (edges) in the GSO. Different from \eqref{topouE:general_problem} in~\cite{segarra2016topoidTSP16}, the formulation \eqref{topouE:general_problem_directed} circumvents computation of eigenvectors (an additional source of errors beyond covariance estimation), and reduces the number of optimization variables.  More importantly, as we show in Section \ref{Ss:Batch_Proximal} it offers favorable structure to invoke a proximal gradient solver with convergence guarantees, even in an online setting where the signals $\bby_t$ arrive in a streaming fashion. 

In closing, we elaborate on $\ccalS$ as the means to enforce the GSO is structurally admissible as well as to incorporate a priori knowledge about $\bbS$. Henceforth we let $\bbS = \bbA$ represent the adjacency matrix of an undirected graph with non-negative weights $(S_{ij}=S_{ji} \geq 0)$ and no self-loops $(S_{ii} = 0)$. Unlike~\cite{segarra2016topoidTSP16,pasdeloup2016inferenceTSIPN16}, we investigate the effect of having  additional prior information about the presence (or absence) of a few edges, or even their corresponding weights. This is well motivated in studies where one can afford to explicitly measure a few of the pairwise relationships among the objects in $\ccalN$. Examples include social network studies involving questionnaire-based random sampling designs~\cite[Ch. 5.3]{kolaczyk2009book}, or experimental testing of suspected regulatory interactions among selected pairs of genes~\cite[Ch. 7.3.4]{kolaczyk2009book}. Since \eqref{topouE:general_problem_directed} is a sparse linear regression problem, one could resort to so-termed variable screening techniques to drop edges prior to solving the optimization; see e.g.,~\cite{screening_camsap17}. Either way, one ends up with extra constraints $S_{ij}\!=\!s_{ij}$, for a few vertex pairs $(i,j)$ in the set $\Omega\subset \ccalN\times \ccalN$ of  observed edge weights $s_{ij}$. Accordingly, we can write the convex set of admissible adjacency matrices as

\begin{align} \label{E:admissible_set_adaptive}
	\ccalS \!:= \! \{ \bbS \, | \, S_{ij}=S_{ji} \geq 0,\: (i,j) \in \ccalN\times \ccalN; \:   S_{ii} = 0, \: i\in \ccalN;\: S_{ij}\! =\! s_{ij},\: (i,j) \in \Omega   \}.
\end{align}
Other GSOs can be accommodated using appropriate modifications.

\subsection{Size of the feasible set} \label{Ss:feasibility_sparse}

Here we examine the feasibility set and the degrees of freedom of the GSO $\bbS$ under the assumption that the perfect output covariance $\bbC_\bby$ is available and $\bbS \in \ccalS$ [cf. \eqref{E:admissible_set_adaptive}]. This would shed light on the dependency of the feasibility set's structure and dimensionality (hence the difficulty of recovering $\bbS$) on the number $|\Omega|$ of observed edges. This dependency can serve as a guideline to the number of edges we may seek to observe prior to graph inference. As we show next, the feasibility set may potentially reduce to a singleton (the graph $\bbS \in \ccalS$ is completely specified by $\bbC_\bby$), or more generally to a low-dimensional subspace. In the latter (more interesting) case, or more pragmatically when we approximate $\hbC_\bby$ with the observations' sample covariance, we solve the convex optimization problem as in~\eqref{topouE:general_problem_directed} to search for a sparse and structurally admissible GSO.
	
Given the (ensemble) output covariance $\bbC_\bby$, consider the mapping $\bbS \bbC_\bby\!=\!\bbC_\bby \bbS$ which stems from the stationarity of $\bby$ (cf. the opening discussion in Section~\ref{Ss:Stationarity_Revisited}). This identity implies that $\bbC_\bby$ and $\bbS$ are simultaneously diagonalizable. Thus the eigenvectors $\bbV$ of $\bbS$ are known and coincide with those of $\bbC_\bby$. Given the GSO eigenvectors $\bbV$, consider the mapping $\bbS\!=\!\bbV \bbLambda \bbV^T$ between $\bbS$ and $\bbLambda$. This can precisely be rewritten as $\text{vec}(\bbS)\!=\!(\bbV \odot \bbV)\bblambda\!=\!\bbW \bblambda$, where $\odot$ denotes the Khatri-Rao (column-wise Kronecker) product, $\bblambda\!\in\!\reals^N$ collects the diagonal entries of $\bbLambda$, and $\bbW\!:=\!\bbV\!\odot\!\bbV\!\in\!\reals^{N^2\!\times\!N}$. Recall that $\bbS \in \ccalS$ and accordingly the entries of $\text{vec}(\bbS)$ corresponding to the diagonal of $\bbS$ are zero. Upon defining the set $\ccalD\!:=\!\big\{N(i-1)\!+\!i \mid i\!\in\!\{1,\ldots,N\}\big\}$, we have the mapping $\bbW_{\ccalD} \bblambda\!=\!\bb0$ to the null diagonal entries of $\bbS$, where $\bbW_{\ccalD}\!\in\!\reals^{N\!\times\!N}$ is a submatrix of $\bbW$ that contains rows indexed by the set $\ccalD$. Thus, it follows $\bbW_{\ccalD}$ is rank-deficient and $\bblambda\in\textrm{ker}(\bbW_{\ccalD})$, where $\textrm{ker}(.)$ denotes the null-space of its matrix argument. In particular, assume that $\text{rank}(\bbW_{\ccalD})\!=\!N\!-\!k$, $1\!\leq\!k\!< N$, which implies $\bblambda$ lives in a $k$-dimensional subspace. As some of the entries in $\bbS$ are known according to $\ccalS$, intuitively we expect that by observing $k=|\Omega|$ ``\textit{sufficiently different}'' edges,  the feasible set may boil down to a singleton resulting in a unique feasible $\bbS \in \ccalS$. To quantify the effect of the partial connectivity constraints on the size of the feasible set, let $\ccalM\!:=\!\{N(j-1)\!+\!i \mid (i,j)\!\in\!\Omega\}$ correspond to the known entries of $\text{vec}(\bbS)$. Then upon defining $\bbU\!\in\!\reals^{N\!\times\!k}$ comprising the basis vectors of $\textrm{ker}(\bbW_{\ccalD})$, the condition $\textrm{rank}(\bbW_{\ccalM} \bbU)\!=\!k$ would be sufficient to determine $\bbS$ uniquely in the $k$-dimensional null space of $\bbW_{\ccalD}$ as stated in the following proposition.

\begin{Proposition}
\label{pro:feasibility}
Suppose the GSO eigenvectors $\bbV$ are given. If $\text{rank}(\bbW_{\ccalD})\!=\!N\!-\!k$ and $\text{rank}(\bbW_{\ccalM} \bbU)\!=\!k$, then $\ccalS$ is a singleton.
\end{Proposition}

\begin{proof}
Since $\bblambda\!\in\!\textrm{ker}(\bbW_{\ccalD})$, there exists an $\bbalpha\!\in\!\reals^k$ such that $\bblambda\!=\bbU \alpha$. From the known entries of $\text{vec}(\bbS)$ denoted by $\bbw\!:=\![\text{vec}(\bbS)]_\ccalM$ we have $\bbW_\ccalM \bblambda\!=\!\bbW_\ccalM \bbU \bbalpha\!=\!\bbw$. Thus, to uniquely identify $\bbalpha$ and equivalently $\bblambda$ (and $\bbS$), it is sufficient to have $\textrm{rank}(\bbW_{\ccalM} \bbU)\!=\!k$.
\end{proof}

\noindent Proposition~\ref{pro:feasibility} further implies that $\textrm{rank}(\bbW_{\ccalM})\!\geq\!k$ under the assumption that $\textrm(\bbW_{\ccalM} \bbU)\!=\!k$. This is due to the inequality $\text{rank}(\bbW_{\ccalM} \bbU) \leq \min\{\textrm{rank}(\bbW_{\ccalM}),\textrm{rank}(\bbU)\}$. Observing $k$ ``\textit{sufficiently different}'' edges for unique recovery of $\bbS$ is the intuition behind the rank constraint on $\bbW_{\ccalM}$. In real-world graphs, we have observed that $k$ is typically much smaller than $N$; see also Section~\ref{S:numerical_results} and \cite[Section~3]{segarra2016topoidTSP16}. This would make it feasible to uniquely identify the graph, given only its eigenvectors and the status of $k$ edges. However, in practice we may not know about the status of those many edges, or, the output covariance $\bbC_\bby$ may only be imperfectly estimated via sample covariance matrix $\hbC_{\bby}$. This  motivates searching for an optimal graph while accounting for the (finite sample size) approximation errors and the prescribed structural constraints, the subject dealt with next.

\subsection{Proximal gradient algorithm for batch topology identification}\label{Ss:Batch_Proximal}

Exploiting the problem structure in \eqref{topouE:general_problem_directed}, a batch proximal gradient (PG) algorithm is developed in this section to recover the network topology; see~\cite{Boyd2014Proximal} for a comprehensive tutorial treatment on proximal methods. Based on this module, an online algorithm for tracking the (possibly dynamically-evolving) GSO from streaming graph signals is obtained in Section \ref{S:Online}.  PG methods have been popularized for $\ell_1$-norm regularized linear regression problems, through the class of iterative shrinkage-thresholding algorithms (ISTA); see e.g.,~\cite{daubechies,wright,beck2017first}. The main advantage of ISTA over off-the-shelf interior point methods is its computational simplicity. We show this desirable feature can permeate naturally to the topology identification context of this paper, addressing the open algorithmic questions in~\cite[Sec. IV]{segarra2016topoidTSP16}. 

To make the graph learning problem amenable to this optimization method, we dualize the constraint $\|\bbS\hbC_{\bby}-\hbC_{\bby} \bbS\|_F^2\leq \epsilon$ in \eqref{topouE:general_problem_directed} and write the composite, non-smooth optimization

\begin{equation}\label{eq:opt_batch} 
		\bbS^{\star} \in \underset{\bbS \in \ccalS}{\text{argmin}} \:
		F(\bbS):= \|\bbS\|_1\!+\!\underbrace{\frac{\mu}{2} \|\bbS \hbC_{\bby}-\hbC_{\bby} \bbS\|_F^2}_{g(\bbS)}.
\end{equation}
The quadratic function $g(\cdot)$ is convex and $M$-smooth [i.e., $\nabla g(\cdot)$ is $M$-Lipschitz continuous] and $\mu>0$ is a tuning parameter. Notice that \eqref{eq:opt_batch} and \eqref{topouE:general_problem_directed} are equivalent optimization problems, since for each $\epsilon$ there exists a value of $\mu$ such that the respective minimizers coincide.

To derive the PG iterations, first notice that the gradient of $g(\bbS)$ in \eqref{eq:opt_batch} has the simple form
\begin{equation} \label{eq:grad_g}
	\nabla g(\bbS) = \mu\big[(\bbS \hbC_{\bby} - \hbC_{\bby} \bbS)\hbC_{\bby} - \hbC_{\bby}(\bbS \hbC_{\bby} - \hbC_{\bby} \bbS)\big],
\end{equation}
which is Lipschitz continuous with constant $M\!=\!4 \mu \lambda_{\max}^{2} (\hbC_{\bby})$. With $\alpha>0$ and $\ccalS$ a convex set, introduce the proximal operator of a function $\alpha f(\cdot):\reals^{N\times N}\to \reals$ evaluated at matrix $\bbM \in \reals^{N \times N}$ as 

\begin{equation} \label{eq:prox}
	\bbZ(\bbM) = \text{prox}_{\alpha f,\ccalS}(\bbM) := \argmin_{\bbX \in \ccalS} \left[f(\bbX) + \frac{1}{2\alpha} \| \bbX -\bbM \|_F^2\right].
\end{equation}
With these definitions, the PG updates with fixed step size $\gamma < \frac{2}{M}$ to solve the batch problem \eqref{eq:opt_batch} are given by ($k=1,2,\ldots$ denote iterations)

\begin{equation} \label{eq:batch_prox}
	\bbS_{k+1} := \text{prox}_{\gamma \| \cdot \|_1,\ccalS}\left(\bbS_k - \gamma \nabla g(\bbS_k) \right).
\end{equation}
It follows that GSO refinements are obtained through the composition of a gradient-descent step and a proximal operator. Instead of directly optimizing the composite cost in \eqref{eq:opt_batch}, the PG update rule in \eqref{eq:batch_prox} can be interpreted as the result of minimizing a quadratic overestimator of $F(\bbS)$ at judiciously chosen points (here the current iterate $\bbS_k$); see~\cite{beck} for a detailed justification. 

Evaluating the proximal operator efficiently is key to the success of PG methods. For our specific case of sparse graph learning with partial connectivity information, i.e., $\ccalS$ as defined in~\eqref{E:admissible_set_adaptive} and $f(\bbS) = \|\bbS\|_1$, the proximal operator $\bbZ$ in \eqref{eq:prox} has entries given by

\begin{equation} \label{eq:prox_S_P} 
	Z_{ij}(M_{ij})=\left\{\begin{array}{cc}
		0, & i=j \\
		s_{ij}, & (i,j) \in \Omega\\
		\text{max}(0, M_{ij} - \alpha), & \text{otherwise}.
	\end{array}\right.
\end{equation}
The resulting entry-wise separable nonlinear map nulls the diagonal entries of $\bbS_{k+1}$, sets the edge weights corresponding to $\Omega$ to the observed values $s_{ij}$, and applies a non-negative soft-thresholding operator to update the remaining entries. Note that for symmetric $\bbS$, the gradient \eqref{eq:grad_g} will also be a symmetric matrix. So if the algorithm is initialized with, say, a very sparse random symmetric matrix, then all subsequent iterates $\bbS_k$, $k\geq 1$, will be symmetric without requiring extra projections. The resulting iterations are tabulated under Algorithm \ref{A:alg1}, which will serve as the basis for the online algorithm in the next section. 

\begin{algorithm}[t!]
    \caption{PG for batch topology identification}
\label{A:alg1}
\algsetup{linenosize=\normalsize}
\begin{algorithmic}[1]
   \REQUIRE  $\hbC_\bby$,  $\mu>0$.\vspace{0.2 cm}
   \STATE Initialize $\bbS_0\neq \mathbf{0}$ as a sparse, random symmetric matrix, $\gamma=1/[4\mu\lambda_{\max}^2(\hbC_\bby)]$, $k=0$.\vspace{0.2 cm} 
   \WHILE {not converged}\vspace{0.1 cm}
   \STATE Compute $\nabla g(\bbS_k) = \mu\big[(\bbS_k \hbC_{\bby} - \hbC_{\bby} \bbS_k)\hbC_{\bby} - \hbC_{\bby}(\bbS_k \hbC_{\bby} - \hbC_{\bby} \bbS_k)\big]$.\vspace{0.2 cm}
   \STATE Take gradient descent step $\bbD_k=\bbS_k-\gamma \nabla g(\bbS_k)$.\vspace{0.2 cm}
   \STATE Update $\bbS_{k+1} = \text{prox}_{\gamma \| \cdot \|_1,\ccalS}\left(\bbD_k \right)$ via the proximal operator in \eqref{eq:prox_S_P}.\vspace{0.2 cm}
   \STATE $k = k + 1$.\vspace{0.2 cm}
   \ENDWHILE \vspace{0.2 cm}
   \RETURN $\bbS^{\star} = \bbS_k$.
\end{algorithmic}
\end{algorithm}

The computational complexity is dominated by the gradient evaluation in \eqref{eq:grad_g}, incurring a cost of $\ccalO(\|\bbS_k\|_0 N^2)$ per iteration $k$. The iterates $\bbS_{k}$ tend to become (and remain) quite sparse at early stages of the algorithm by virtue of the soft-thresholding operations (a sparse initialization is useful to this end), hence reducing the complexity of the matrix products in \eqref{eq:grad_g}. As $k\to\infty$ the sequence of iterates \eqref{eq:batch_prox} provably converges to a minimizer $\bbS^{\star}$ [cf. \eqref{eq:opt_batch}]; see e.g.,~\cite{bauschke2011convex}. Moreover, $F(\bbS_k) - F(\bbS^{\star}) \rightarrow 0$ due to the continuity of the composite objective function $F(\cdot)$ -- a remark on the convergence rate is now in order.

\begin{Remark}
Results in~\cite{nesterov2} assert that convergence speedups can be obtained through the so-termed accelerated (A)PG algorithm; see~\cite{beck} for specifics in the context of ISTA that are also applicable here. Without increasing the computational complexity of Algorithm \ref{A:alg1}, one can readily devise an accelerated variant with a (worst-case) convergence rate guarantee of $\ccalO(1/\sqrt{\varepsilon})$ iterations to return an $\varepsilon$-optimal solution measured by the objective value $F(\cdot)$ [cf. Algorithm \ref{A:alg1} instead offering a $\ccalO(1/\varepsilon)$ rate].     
\end{Remark} 

\section{Online network topology inference}\label{S:Online}

Additional challenges arise with real-time network data collection, where analytics must often be performed ``on-the-fly'' and without the opportunity to revisit past graph signal observations due to e.g., staleness or storage constraints~\cite{kostas2014spmag}. Consider now \emph{online} estimation of $\bbS$ (or even tracking $\bbS_t$ in a dynamic setting) from streaming data $\{\bby_1,\ldots, \bby_t, \bby_{t+1},\ldots \}$. To this end, a viable approach is to solve at each time instant $t=1,2,\ldots,$ the composite, time-varying optimization problem [cf. \eqref{eq:opt_batch}]
\begin{equation}\label{eq:opt_online}
		\bbS^{\star}_t \in \underset{\bbS \in \ccalS}{\text{argmin}} \:
		F_t(\bbS) := \|\bbS\|_1\!+\!\underbrace{\frac{\mu}{2} \|\bbS \hbC_{\bby,t}-\hbC_{\bby,t} \bbS\|_F^2}_{g_t(\bbS)}.
\end{equation}
In writing $\hbC_{\bby,t}$ we make explicit that the covariance matrix is estimated with all signals acquired by time $t$. As data come in, the covariance estimate will fluctuate and hence the time dependence of the objective function $F_t(\bbS)$ through its smooth component $g_t$. Notice that even as $t$ becomes large, the squared residuals in $g_t$ remain roughly of the same order due to data averaging (rather than accumulation) in $\hbC_{\bby,t}$. Thus a constant regularization parameter $\mu>0$ tends to suffice because $g_t(\bbS)$ will not dwarf the $\ell_1$-norm term in \eqref{eq:opt_online}.

The solution $\bbS^{\star}_t$ of \eqref{eq:opt_online} is the batch network estimate at time $t$. Accordingly, a naive sequential estimation approach consists of solving \eqref{eq:opt_online} using Algorithm \ref{A:alg1} repeatedly. However online operation in delay-sensitive applications may not tolerate running multiple inner PG iterations per time interval, so that convergence to $\bbS^{\star}_t$  is attained for each $t$ as required by Algorithm \ref{A:alg1}. If $\ccalG$ is dynamic it may not be even prudent to obtain $\bbS^{\star}_t$ with high precision (hence incurring high delay and unnecessary computational cost), since at time $t+1$ a new datum arrives and the solution $\bbS^{\star}_{t+1}$ may deviate significantly from the prior estimate. These reasons motivate devising an efficient online and recursive algorithm to solve the time-varying optimization problem \eqref{eq:opt_online}. We are faced with an interesting trade-off that emerges with time-constrained data-intensive problems, where a high-quality answer that is obtained slowly can be less useful than a medium-quality answer that is obtained quickly. 

\subsection{Algorithm construction}\label{Ss:algo_construction}

Our algorithm construction approach entails two steps per time instant $t=1, 2, \ldots$,  where we: (i) recursively update the observations' covariance matrix $\hbC_{\bby,t}$ in $\mathcal{O}(N^2)$ complexity; and then (ii) run a single iteration of the batch graph learning algorithm developed in Section \ref{Ss:Batch_Proximal} to solve \eqref{eq:opt_online} efficiently. Different from recent approaches that learn dynamic graphs from the observation of smooth signals~\cite{kalofolias2017icassp,cardoso2020financial}, the resulting algorithm's memory storage requirement and computational cost per data sample $\bby_t$ does not grow with $t$.  A similar idea to the one outlined in (ii) was advocated in~\cite{Liam_online} to develop an online sparse subspace clustering algorithm; see also~\cite{BainganaInfoNetworks} for dynamic SEM estimation from traces of information cascades.

Step (i) is straightforward, and the sample covariance $\hbC_{\bby,t-1}$ is recursively updated once $\bby_{t}$ becomes available through a rank-one correction as follows

\begin{equation}\label{E:update_covariance}
	\hbC_{\bby,t} = \frac{1}{t}\left((t-1)\hbC_{\bby,t-1}+\bby_{t}\bby_{t}^{T}\right).
\end{equation}
This so-termed infinite memory sample estimate is appropriate for time-invariant settings, i.e., when the graph topology remains fixed for all $t$. To track dynamic graphs, it is prudent to eventually forget about past observations [cf. \eqref{E:update_covariance} incorporates all signals $\{\bby_\tau\}_{\tau=1}^t$]. This can be accomplished via exponentially-weighted empirical covariance estimators or by using a (fixed-length) sliding window of data, both of which can be updated recursively via minor modifications to \eqref{E:update_covariance}. Initialization of the covariance estimate $\hbC_{\bby,0}$ can be performed in practice via sample averaging of a few signals collected before the algorithm runs, or simply as a scaled identity matrix.

To solve \eqref{eq:opt_online} online by building on the insights gained from the batch solver [cf. step (ii)], we let iterations $k=1,2,\ldots$ in Algorithm \ref{A:alg1} coincide with the instants $t$ of data acquisition. In other words, at time $t$ we run \emph{a single iteration} of \eqref{eq:batch_prox} to update $\bbS_{t}$ before the new datum $\bby_{t+1}$ arrives at time $t+1$. Specifically, the online PG algorithm takes the form

\begin{equation} \label{eq:batch_prox_online}
	\bbS_{t+1} := \text{prox}_{\gamma_t \lVert \cdot \rVert_1,\ccalS}\big(\bbS_t - \gamma_t \nabla g_t(\bbS_t) \big),
\end{equation}
where the step size $\gamma_t$ is chosen such that $\gamma_t < \frac{2}{M_t} = \frac{1}{2 \mu  \lambda_{\max}^{2} (\hbC_{\bby,t})}$. Recall that the gradient $\nabla g_t(\bbS_t)$ is given by \eqref{eq:grad_g}, and it is a function of the updated covariance matrix $\hbC_{\bby,t}$ [cf. \eqref{E:update_covariance}]. The proximal operator for the $\ell_1-$norm entails the pointwise nonlinearity in \eqref{eq:prox_S_P}, with a threshold $\gamma_t$ that will in general be time varying. If the signals arrive faster, one can create a buffer and perform each iteration of the algorithm on a $\hbC_{\bby,t}$ updated with a sliding window of all newly observed signals (in a way akin to processing a minibatch of data). On the other hand, for a slower arrival rate additional PG iterations during $(t-1,t]$ would likely improve recovery performance; see also the discussion following Theorem \ref{theorem:str_conv}. The proposed online scheme is tabulated under Algorithm \ref{A:alg2}; the update of $\hbC_{\bby,t}$ can be modified to accommodate tracking of dynamic graph topologies.

\begin{algorithm}[t!]
    \caption{PG for online topology identification}
\label{A:alg2}
\algsetup{linenosize=\normalsize}
\begin{algorithmic}[1]
   \REQUIRE  $\{\bby_1,\ldots, \bby_t, \bby_{t+1},\ldots \}$, $\hbC_{\bby,0}$,  $\mu>0$.\vspace{0.2 cm}
   \STATE Initialize $\bbS_1\neq \mathbf{0}$ as a sparse, random symmetric matrix, $\gamma_0=1/[4\mu\lambda_{\max}^2(\hbC_{\bby,0})]$, $k=0$.\vspace{0.2 cm} 
   \FOR{$t = 1, 2, \dots$}\vspace{0.1 cm}
   \STATE Update $\hbC_{\bby,t} = \frac{1}{t}\left((t-1)\hbC_{\bby,t-1}+\bby_{t}\bby_{t}^{T}\right)$ and $\gamma_t$.\vspace{0.2 cm}
   \STATE Compute $\nabla g_t(\bbS_t) = \mu\big[(\bbS_t \hbC_{\bby,t} - \hbC_{\bby,t} \bbS_t)\hbC_{\bby,t} - \hbC_{\bby,t}(\bbS_t \hbC_{\bby,t} - \hbC_{\bby,t} \bbS_t)\big]$.\vspace{0.2 cm}
   \STATE Take gradient descent step $\bbD_t=\bbS_t-\gamma_t \nabla g_t(\bbS_t)$.\vspace{0.2 cm}
   \STATE Update $\bbS_{t+1} = \text{prox}_{\gamma_t \| \cdot \|_1,\ccalS}\left(\bbD_t \right)$ via the proximal operator in \eqref{eq:prox_S_P}.\vspace{0.2 cm}
   \ENDFOR \vspace{0.2 cm}
   \RETURN $\bbS_{t+1}$.
\end{algorithmic}
\end{algorithm}

\begin{Remark}\label{R:comparisons_conference}
In conference precursors to this paper~\cite{DSWonlinetopoid,RG_topoID_camsap_online}, different algorithms were developed to track graph topologies from streaming stationary signals. The ADMM-based scheme in~\cite{DSWonlinetopoid} minimizes an online criterion stemming from \eqref{topouE:general_problem} and hence one needs to track sample covariance eigenvectors; the same is true for the online alternating-minimization algorithm in~\cite{RG_topoID_camsap_online}. The merits of working with the inverse problem \eqref{topouE:general_problem_directed} are outlined in Section \ref{Ss:Stationarity_Revisited}. Moreover, enforcing some of the admissibility constraints in $\ccalS$ requires an inner ADMM loop to compute nontrivial projection operators, which could hinder applicability in delay-sensitive environments~\cite{DSWonlinetopoid}. Neither of these schemes offer convergence guarantees to the solutions of the resulting time-varying optimization problems. For Algorithm \ref{A:alg2}, these type of results are established in the ensuing section. 
\end{Remark} 

\subsection{Convergence and regret analysis}\label{Ss:Convergence}

The key difference between the batch algorithm \eqref{eq:batch_prox} and its online counterpart \eqref{eq:batch_prox_online} is the variability of $g_{t}$ per iteration in the latter. Ideally, we would like Algorithm \ref{A:alg2} to closely track the sequence of minimizers $\left\{\bbS_{t}^{\star}\right\}$ for large enough $t$, something we corroborate numerically in Section~\ref{S:numerical_results}. Following closely the analysis in~\cite{Liam_online}, we derive recovery (i.e., tracking error) bounds $\left\|\bbS_{t}-\bbS_{t}^{\star}\right\|_F$ under the pragmatic assumption that $g_t$ is strongly convex and $\bbS_{t}^{\star}$ is the unique minimizer of \eqref{eq:opt_online}, for each $t$. Before stating the main result in Theorem \ref{theorem:str_conv}, the following proposition (proved in Appendix \ref{App:proof_strong_convx}) offers a condition for strong convexity of $g_t$.

\begin{Proposition}
\label{pro:strong_convx}
Let set $\ccalD$ contain the indices of $\text{vec}(\bbS)$ corresponding to the diagonal entries of $\bbS$; i.e., $\ccalD\!:=\!\big\{N(i-1)\!+\!i \mid i\!\in\!\{1,\ldots,N\}\big\}$, and let $\ccalD^{c}$ be the complement of $\ccalD$. Define $\bbPsi_{t} =\hbC_{\bby,t} \otimes \bbI_N-\bbI_N \otimes \hbC_{\bby,t}\in \reals^{N^2\times N^2}$. If $\bbPsi_{t, \ccalD^{c}}\in \reals^{N^2\times N(N-1)}$ (the submatrix of $\bbPsi_{t}$ that contains columns indexed by the set $\ccalD^{c}$) is full column rank, then $g_t(\bbS)$ in \eqref{eq:opt_online} is strongly convex with constant $m_t>0$ being the smallest (nonzero) singular value of $\bbPsi_{t, \ccalD^{c}}$.
\end{Proposition}
Consider the eigendecomposition $\hbC_{\bby,t}=\hbV_t\hbLambda_t\hbV_t^T$ of the sample covariance matrix at time $t$, with $\hbV_t$ denoting the matrix of orthogonal eigenvectors and $\hbLambda_t$ the diagonal matrix of non-negative eigenvalues. Exploiting the structure of $\bbPsi_{t}$ it follows that the strong convexity condition stated in Proposition \ref{pro:strong_convx} will be satisfied if (i) all eigenvalues of $\hbC_{\bby,t}$ are distinct; and (ii) the matrix $\hbV_t\circ \hbV_t$ is non-singular. Recalling the covariance expression in \eqref{eqn_diagonalize_covariance_adaptive}, the aforementioned conditions (i)-(ii) immediately translate to properties of the diffusion filter's (squared) frequency response and the GSO eigenvectors. In extensive simulations involving several synthetic and real-world graphs, we have indeed observed that $\bbPsi_{t, \ccalD^{c}}$ is typically full column rank and thus $g_t$ is strongly convex. 

Under the strong convexity assumption, we have the following (non-asymptotic) performance guarantee for Algorithm \ref{A:alg2}. The result is adapted from~\cite[Theorem 1]{Liam_online}; see Appendix \ref{App:proof_str_conv} for a proof.
\begin{Theorem} \label{theorem:str_conv}
Let $\nu_{t}:=\!\left\|\bbS_{t+1}^{\star}\!-\!\bbS_{t}^{\star}\right\|_{F}$ capture the variability of the optimal solution of \eqref{eq:opt_online}. If $g_t$ in \eqref{eq:opt_online} is strongly convex with constant $m_t$, then for all $t\geq1$ the iterates $\bbS_t$ generated by Algorithm \ref{A:alg2} satisfy
	
\begin{equation} \label{eq:theorem_cvx_case}
\left\|\bbS_{t}-\bbS_{t}^{\star}\right\|_F \leq \tilde{L}_{t-1}\left(\left\|\bbS_{0}-\bbS_{0}^{*}\right\|_F+\sum_{\tau=0}^{t-1} \frac{\nu_{\tau}}{\tilde{L}_{\tau}}\right),
\end{equation}
where $L_{t}=\max \left\{\left|1-\gamma_{t} m_{t}\right|,\left|1-\gamma_{t} M_{t}\right|\right\}, \tilde{L}_{t}=\prod_{\tau=0}^{t} L_{\tau}$. In terms of the objective values, it can be shown that $F_{t}\left(\bbS_{t}\right)-F_{t}\left(\bbS_{t}^{\star}\right) \leq \frac{M_{t}}{2}\left\|\bbS_{t}-\bbS_{t}^{\star}\right\|_F^{2}$; see~\cite[Theorem 10.29]{beck2017first}.
\end{Theorem}
As expected, Theorem~\ref{theorem:str_conv} asserts that the higher the variability in the underlying graph, the higher the recovery performance penalty. Even if the graph $\ccalG$ (and hence the GSO) is time invariant, then $\nu_t$ will be non-zero especially for small $t$ since the solution $\bbS_{t}^{*}$ may fluctuate due to insufficient data. Much alike classic performance results in adaptive signal processing~\cite{Solo_Adaptive_Book}, here there is misadjustment due to the ``dynamics-induced noise'' $\nu_{t}$ and hence the online algorithm will only converge to within a neighborhood of $\bbS_{t}^{\star}$.

To better distill what determines the size of the convergence radius, define $\hat{L}_{t}:= \max_{\tau=0,\ldots,t} L_{\tau}$, $\hat{\nu}_{t}:=\max_{\tau=0,\ldots,t} \nu_{\tau}$ and sum the geometric series in the right-hand side of \eqref{eq:theorem_cvx_case} to obtain the simplified bound

\begin{equation} \label{eq:theorem_cvx_case_simlpified}
 \left\|\bbS_{t}-\bbS_{t}^{\star}\right\|_F \leq\left(\hat{L}_{t-1}\right)^{t}\left\|\bbS_{0}-\bbS_{0}^{\star}\right\|_F+\frac{\hat{\nu}_{t}}{1-\hat{L}_{t-1}}.
\end{equation}
Further suppose $m_{\tau} \geq m$ and $M_{\tau} \leq M$ and the step-size chosen as $\gamma_{\tau}=2 /(m_{\tau}+M_{\tau})$, for all $\tau=0, \ldots, t,$. Then it follows that $\hat{L}_{t} \leq(M-m) /(M+m)<1$. The misadjustment $\hat{\nu}_{t}/(1-\hat{L}_{t-1})$ in \eqref{eq:theorem_cvx_case_simlpified} grows with $\hat{\nu}_{t}$ as expected, and will also increase when the problem is badly conditioned ($M\to \infty$ or $m\to 0$) because $\hat{L}_{t}\to 1$. If one could afford taking $i_t$ PG iterations (instead of a single one as in Algorithm \ref{A:alg2}) per time step $t$, the performance gains can be readily evaluated by substituting $\tilde{L}_{t}=\prod_{\tau=0}^{t} L_{\tau}^{i_{\tau}}$ in Theorem~\ref{theorem:str_conv} to use the bound \eqref{eq:theorem_cvx_case}. 


In closing, we state dynamic regret bounds which are weaker than the performance guarantees in \eqref{eq:theorem_cvx_case}-\eqref{eq:theorem_cvx_case_simlpified}, but they do not require strong convexity assumptions. The proof of the following result can be found in~\cite{Liam_online} and is omitted here in the interest of brevity. 
 
\begin{Theorem}\label{theorem:regret_bound}
Let $\hat{\bbS}_{t}^{\star}\!=\!\frac{1}{t} \sum_{\tau=0}^{t-1} \bbS_{\tau}^{\star}$ be the average trajectory of the optimal solutions of \eqref{eq:opt_online}, and introduce $\rho_{t}(\tau):=\!\left\|\hat{\bbS}_{t}^{\star}-\bbS_{\tau}^{\star}\right\|$ to capture the deviation of instantaneous solutions from this average trajectory. Define $\delta_{t}:=\!\left\|g_{t+1}-g_{t}\right\|_{\infty}$ as a measure of the variability in $F_t$. Suppose $M_{\tau}\leq M$ and set the step-size $\gamma_{\tau}\!=\!\frac{1}{M}$, for all $\tau=0,\ldots,t$. Then for all $t\geq1$ the iterates $\bbS_t$ generated by Algorithm \ref{A:alg2} satisfy
 
\begin{align}\label{eq:regret_bound}
 \frac{1}{t} \sum_{\tau=0}^{t-1}\left(F_{\tau}\left(\bbS_{\tau}\right)-F_{\tau}\left(\bbS_{\tau}^{\star}\right)\right) \leq{} &\frac{M}{2 t}\left\|\bbS_{0}-\hat{\bbS}_{t}^{*}\right\|_F^{2}\nonumber\\
 {}&+\frac{1}{t}\left[F_{0}\left(\bbS_{0}\right)
-F_{t-1}\left(\bbS_{t}\right)+\sum_{\tau=0}^{t-2} \delta_{\tau}+\sum_{\tau=0}^{t-1} \rho_{t}(\tau)\left(N+\frac{M}{2} \rho_{t}(\tau)\right)\right].
\end{align}
\end{Theorem}
To interpret the result of Theorem \ref{theorem:regret_bound}, define $\hat{\rho}_{t}:= \max_{\tau=0,\ldots, t-1}  \rho_{t}(\tau)$ and
$\hat{\delta}_{t}:= \max_{\tau=0, \ldots, t-2} \delta_{\tau}.$ Using these definitions, the following simplified regret bound is obtained immediately from \eqref{eq:regret_bound}

\begin{equation}
\frac{1}{t} \sum_{\tau=0}^{t-1}\left(F_{\tau}(\bbS_{\tau})-F_{\tau}(\bbS_{\tau}^{\star})\right) \leq \frac{1}{t}\left[\frac{M}{2}\left\|\bbS_{0}-\hbS_{t}^{\star}\right\|^{2}
 +F_{0}(\bbS_{0})-F_{t-1}(\bbS_{t})\right]+\frac{M \hat{\rho}_{t}^{2}}{2}+N \hat{\rho}_{t}+\hat{\delta}_{t}.
\end{equation}
If $\hat{\delta}_{t}$ and $\hat{\rho}_{t}$ are well-behaved (i.e., the cost function changes slowly and so does its optimal solution $\bbS_{t}^{*}$) then, on average, $F_{t}\left(\bbS_{t}\right)$ hovers within a constant term of $F_{t}\left(\bbS_{t}^{*}\right)$. The network size $N$ and the problem conditioning (as measured by the Lipschitz constant upper bound $M$) also play a natural role.


\section{Numerical Tests}\label{S:numerical_results}

Here we assess the performance of the proposed algorithms in recovering sparse real-world graphs. To that end, we: (i)
illustrate the scalability of Algorithm~\ref{A:alg1} using a relatively large-scale Facebook
graph with thousands of nodes in a batch setup; (ii) evaluate the performance of the proposed online Algorithm~\ref{A:alg2}
in settings with streaming signals; and (iii) demonstrate the effectiveness of Algorithm~\ref{A:alg2} in adapting to dynamical behavior of the network.

Throughout this section, we infer unweighted real-world networks from the observation of diffusion processes that are
synthetically generated via graph filtering as in \eqref{E:Filter_input_output_time}. For the graph shift $\bbS\!=\!\bbA$, the adjacency matrix of the sought network, we consider a second-order filter $\bbH\!=\!\sum_{l=0}^{2} h_{l} \bbS^{l}$, where the coefficients $\{h_{l}\}$ are drawn uniformly from $[0,1]$. To measure the edge-support recovery, we compute the F-measure defined as the harmonic mean of edge precision and recall (precision is the percentage of correct edges in $\hbS$, and recall is the fraction of edges in $\bbS$ that are retrieved).
	
\subsection{Facebook friendship graph: Offline}\label{Ss:Facebook} 

To evaluate the scalability of Algorithm~\ref{A:alg1}, consider a directed network of $N\!=\!2888$ Facebook users, where the $2981$ edges represent friendships among the users \cite{konect,leskovec2012learning}. More precisely, an edge from node $i$ to node $j$ exists if user $i$ is a friend of the user $j$. To make the graph amenable to our framework, we assume that the friendships are bilateral and ignore the directions. First, we notice that $\textrm{rank}(\bbW_{\ccalD})=2882$ (cf.~Proposition~\ref{pro:feasibility}). This means that knowing the GSO's eigenvectors and $k\!=\!6$ suitably chosen edges as a priori information would lead to a singleton feasibility set. To test Algorithm~\ref{A:alg1} in recovering this reasonably large-scale graph, the performance is averaged over $10$ experiments wherein we assume that we know the existence of $|\Omega|=5$ randomly sampled edges in each experiment. We then generate $T=10^3$, $10^4$, $10^5$, or $10^6$ synthetic random signals $\{\bby_{t}\}_{t=1}^T$ adhering to generative diffusion process in \eqref{E:Filter_input_output_time}, where the entries of the inputs $\{\bbx_{t}\}_{t=1}^T$ are drawn independently from the standard Gaussian distribution yo yield stationary observations. We obtain $\hbC_\bby$ via a sample covariance estimate. For the aforementioned different values of $T$, the recovered $\hbS$ obtained after running Algorithm~\ref{A:alg1} results in the following F-measures. 

\begin{center}
 \begin{tabular}{c | c c c c} 
 Number of observations $(T)$ & $10^3$ & $10^4$ & $10^5$ & $10^6$ \\ [1ex] 
 \hline \hline
 F-measure & $0.45$ & $0.77$ & $0.87$ & $0.94$ \\ [1ex]
\end{tabular}
\end{center}
As the number of observations increase, the estimate $\hbC_\bby$ becomes more reliable which leads to a better performance in recovering the underlying GSO; recall that perfect support recovery corresponds to an F-measure of $1$. Moreover, the reported results corroborate the effectiveness of Algorithm~\ref{A:alg1} in recovering large-scale graphs. Off-the-shelf algorithms utilized to solve related topology inference problems in~\cite{segarra2016topoidTSP16} are not effective for graphs with more than a few hundred nodes.

\subsection{Zachary's karate club: Online}\label{Ss:Zachary} 

Next, we consider the social network of Zachary's karate club~\cite[Ch. 1.2.2]{kolaczyk2009book} represented by a graph $\ccalG$ consisting of $N=34$ nodes or members of the club and $78$ undirected edges symbolizing friendships among them. We seek to infer this graph from the observation of diffusion processes that are synthetically generated via graph filtering as in \eqref{E:Filter_input_output_time}. The rank of $\bbW_{\ccalD}$ (cf. Proposition~\ref{pro:feasibility}) for this graph is $32$. This implies that the knowledge of the perfect output covariance $\bbC_\bby$  leaves the GSO $\bbS$ in a $2$-dimensional subspace which can lead to a singleton feasibility set by observing only $2$ different edges. However, here we work with a noisy sample covariance $\hbC_\bby$. We observe one of the $78$ edges (chosen uniformly at random) and aim to infer the rest of the edges. At each time step, $10$ synthetic signals $\{\bby_p\}$ are generated through diffusion process $\bbH$ where the entries of the inputs $\{\bbx_p\}$ are drawn independently from the standard Gaussian distribution. In the online case, upon sensing $10$ signals at each time step, we first update the sample covariance $\hbC_{\bby,t}$ and then carry out $i_t=10$ PG iterations as Algorithm~\ref{A:alg2}. Also, to examine the tracking capability of the online estimator, after $5000$ time steps, we remove $10\%$ of the existing edges and add the same number of edges elsewhere. This would affect the graph filter $\bbH$ accordingly.

\begin{figure*}[t] 
	\begin{minipage}[b]{0.48\textwidth} 
		\centering
		\includegraphics[width=1\linewidth]{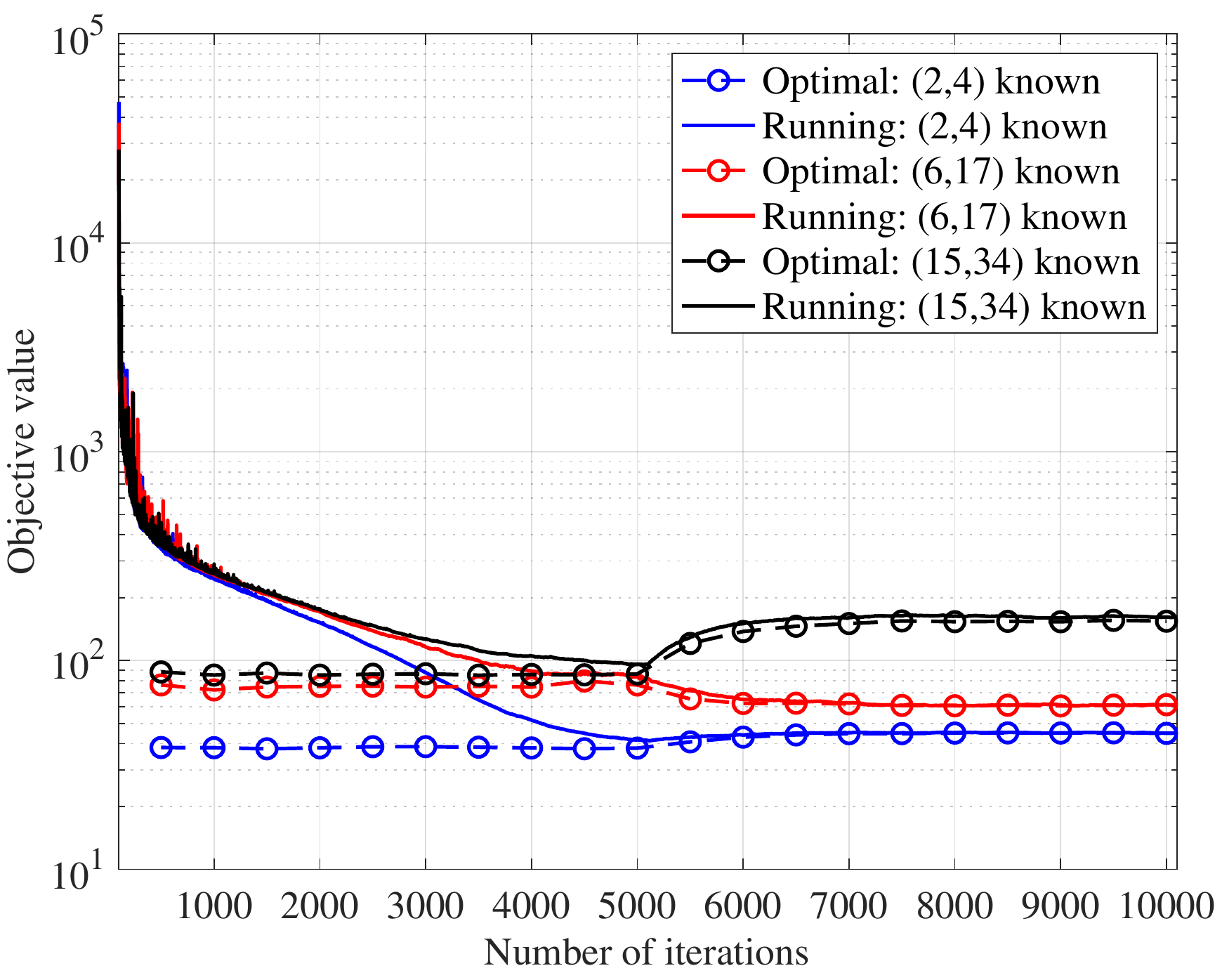}
		\centerline{(a)}\medskip
	\end{minipage}
	\hfill
	\begin{minipage}[b]{0.48\textwidth} 
		\centering
		\includegraphics[width=1\linewidth]{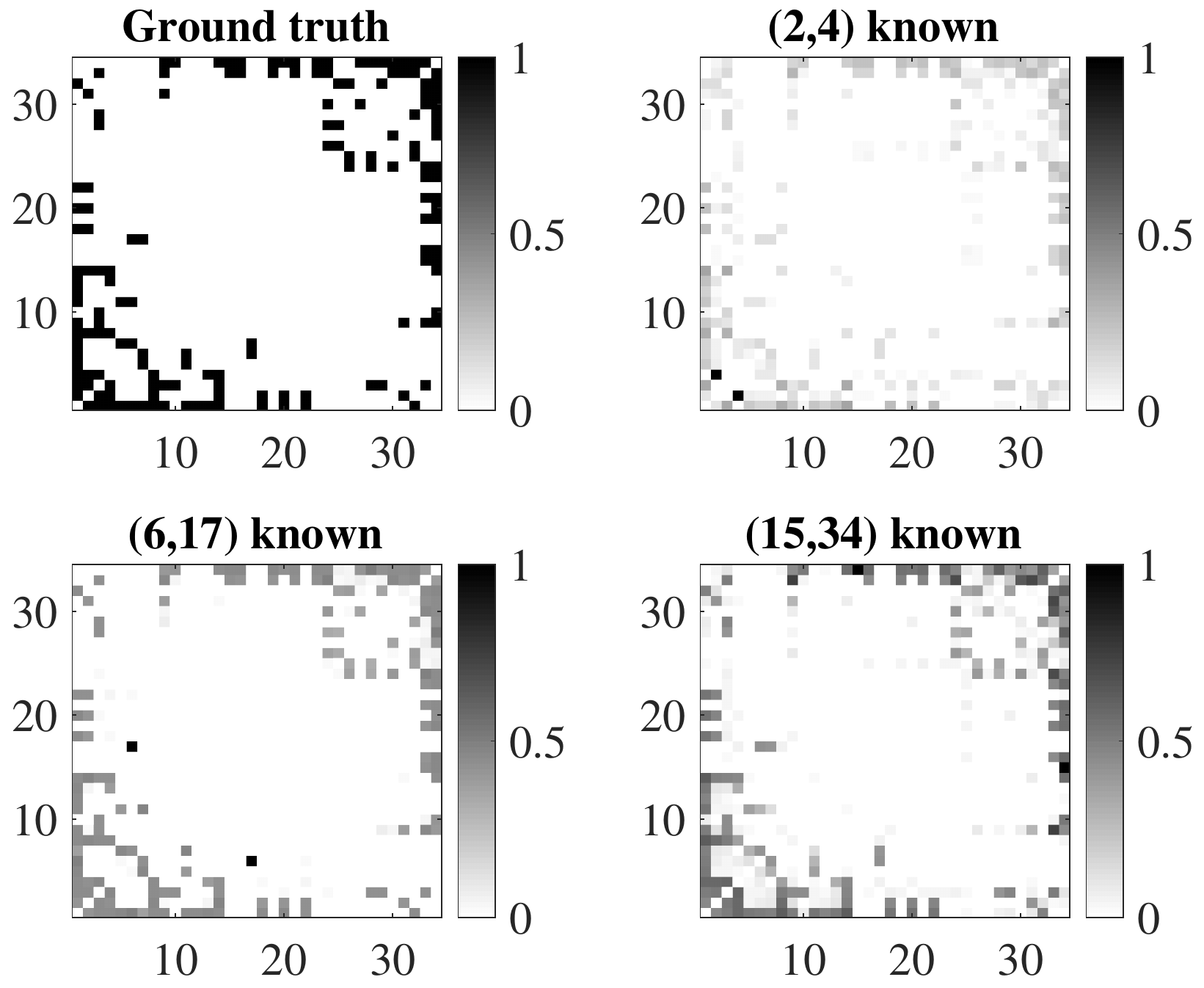}
		\centerline{(b)}\medskip
	\end{minipage}
	\caption{Recovery of Zachary's karate club graph with $N\!=\!34$ nodes. (a) Evolution of the objective values for the online and batch estimators in inferring a karate club. (b) Ground truth adjacency matrix and corresponding estimates with different a priori information on the connectivities attained after $5000$ time steps. 
	}
\end{figure*}

To corroborate the assumption in Theorem~\ref{theorem:str_conv}, it is worth mentioning that throughout the process we observed that $\bbPsi_{t, \ccalD^{c}}$ was full column rank and thus the cost in \eqref{eq:opt_online} was strongly convex; see Proposition~\ref{pro:strong_convx}. Fig.~1-(a) depicts the running objective value $F_t(\bbS_t)$ [cf.~\eqref{eq:opt_online}] averaged over $10$ experiments as a function of the time steps and the a priori knowledge -- $3$ randomly chosen edges. We also superimpose Fig.~1-(a) with the optimal objective value  $F_t(\bbS_t^{\star})$ at each time step. First, we notice that the objective value trajectory converges to a region above the optimal trajectory. Also, we observe that after $5000$ iterations, the performance deteriorates at first due to the sudden change of the network structure, but after observing enough new samples Algorithm \ref{A:alg2} can adapt and track the batch estimator as well. This demonstrates the effectiveness of the developed online algorithm when it comes to adapting to network perturbations.

Finally, we study the quality of the online learned graph $\bbS_t$ at iteration $5000$. Fig.~1-(b)
depicts the heat maps of the ground-truth and inferred adjacency matrices for different a priori information. Although the procedure results in a slight gap between $F_t(\bbS_t^{\star})$ and $F_t(\bbS_t)$, it still reveals the underlying support of $\bbA$ with reasonable accuracy. Interestingly, we notice that an edge with lower betweenness centrality [e.g., $(6,17)$ and $(15,34)$ compared to $(2,4)$] is more informative to identify the network topology; see also \cite{RG_topoID_camsap_online}.

\section{Conclusions}\label{S:conclusions}

We studied the problem of identifying the  topology of an undirected network from streaming observations of stationary signals diffused on the unknown graph. This is an important problem, because as a general principle network structural information can be used to understand, predict, and control the behavior of complex systems. The stationarity assumption implies that the observations' covariance matrix and the GSO commute under mild requirements. This motivates formulating the topology inference task as an inverse problem, whereby one searches for a (e.g., sparse) GSO that is structurally admissible and approximately commutes with the observations' empirical covariance matrix. For streaming data said covariance can be updated recursively, and we show online proximal gradient iterations can be brought to bear to efficiently track the time-varying solution of the inverse problem with quantifiable recovery guarantees. Ongoing work includes extensions of the online graph learning framework to observations of streaming signals that are smooth on the sought network.


\funding{Work in this paper was supported by the National Science Foundation awards CCF-1750428 and ECCS-1809356.}

\appendixtitles{yes} 
\appendix
\section{Proof of Proposition \ref{pro:strong_convx}}\label{App:proof_strong_convx}

Strong convexity of $g_t(\bbS)$ in \eqref{eq:opt_online} is equivalent to finding $m_{t}>0$ such that

\begin{equation} \label{eq:cond:str:cvx}
	\langle \bbS_1-\bbS_2, \nabla g_t(\bbS_1)-\nabla g_t(\bbS_2) \rangle \geq m_{t} \| \bbS_1-\bbS_2\|_{F}^{2},
\end{equation}
for $\bbS_1$, $\bbS_2 \in \ccalS$. Note that the Frobenius inner product of two real square matrices is defined as $\langle\mathbf{A}, \mathbf{B}\rangle=\sum_{i, j} A_{i j} B_{i j}=\operatorname{trace}(\mathbf{A}^{T} \mathbf{B})$. Substituting the time-varying counterpart of \eqref{eq:grad_g} in \eqref{eq:cond:str:cvx} and
using properties of the Khatri-Rao product, one can write the left hand side (LHS) of \eqref{eq:cond:str:cvx} as

\begin{equation*} 
		\| (\bbS_1 - \bbS_2) \hbC_{\bby,t} - \hbC_{\bby,t} (\bbS_1 - \bbS_2) \|_{F}^{2} = \| \bbPsi_t \text{vec}(\bbS_1 - \bbS_2) \|_2^2,
\end{equation*}
where $\text{vec}(\cdot)$ stands for matrix vectorization and $\| \cdot \|$ denotes the Euclidean distance.
Since GSOs are devoid of self-loops, we can discard the zero entries of $\text{vec}(\bbS_1 - \bbS_2)$ corresponding to the diagonal entries of $\bbS$ as well as the related columns in $\bbPsi_t$. This would further simplify the LHS of \eqref{eq:cond:str:cvx} leading immediately to the result in Proposition \ref{pro:strong_convx}. Also, notice that $\| \text{vec}(\bbS_1-\bbS_2)\|_{2}^{2} = \| \bbS_1-\bbS_2\|_{F}^{2}$. 

\section{Proof of Theorem \ref{theorem:str_conv}}\label{App:proof_str_conv}

For any $\bbS, \bbS^{\prime} \in \ccalS$, we have

\begin{align} \label{eq:L_t}
		\left\|\bbS-\gamma_{t} \nabla g_{t}(\bbS)-\left[\bbS^{\prime}-\gamma_{t} \nabla g_{t}\left(\bbS^{\prime}\right)\right]\right\|_F^{2} 
		&{}\leq\left(1-2 \gamma_{t} m_{t}+\gamma_{t}^{2} M_{t}^{2}\right)\left\|\bbS-\bbS^{\prime}\right\|_F^{2}\nonumber \\
		&{}\leq L_{t}^{2}\left\|\bbS-\bbS^{\prime}\right\|_F^{2},
\end{align}
where the first inequality is due to the Lipschitz continuity of $g_t(\cdot)$ with constant $M_t\!=\!4 \mu \lambda_{\max}^{2} (\hbC_{\bby,t})$ along with the strong convexity condition \eqref{eq:cond:str:cvx}. The second inequality holds since $M_t \geq m_t$.

Noting that $\bbS_{t}^{*}$ is a fixed point of \eqref{eq:batch_prox_online}, we have

\begin{align*}
\left\|\bbS_{t+1}-\bbS_{t}^{*}\right\|_F & {}= \big\lVert \text{prox}_{\gamma_t \lVert \cdot \rVert_1,\ccalS}\big(\bbS_t  - \gamma_t \nabla g_t(\bbS_t) \big) - \text{prox}_{\gamma_t \lVert \cdot \rVert_1,\ccalS}\big(\bbS_{t}^{*} - \gamma_t \nabla g_t(\bbS_{t}^{*}) \big) \big\rVert_F \\
& {}\leq \left\| \bbS_t  - \gamma_t \nabla g_t(\bbS_t)  -\left[\bbS_{t}^{*} - \gamma_t \nabla g_t(\bbS_{t}^{*}) \right] \right\|_F \\
& {} \leq L_{t}\left\|\bbS_t-\bbS_{t}^{*}\right\|_F,
\end{align*}
where the first inequality is due to the nonexpansiveness of the proximal operator and the second one follows from \eqref{eq:L_t}.

Leveraging the triangle inequality and the definition $\nu_{t}\!=\!\left\|\bbS_{t+1}^{*}\!-\!\bbS_{t}^{*}\right\|_{F}$ results in 

\begin{align*}
\left\|\bbS_{t+1}-\bbS_{t+1}^{*}\right\|_F &{} \leq\left\|\bbS_{t+1}-\bbS_{t}^{*}\right\|_F+\left\|\bbS_{t+1}^{*}-\bbS_{t}^{*}\right\|_F \\
&{} \leq L_{t}\left\|\bbS_{t}-\bbS_{t}^{*}\right\|_F+\nu_{t},
\end{align*}
which can be applied recursively to obtain \eqref{eq:theorem_cvx_case}. 

%
%
%
%

\reftitle{References}


\externalbibliography{yes}
\bibliography{your_external_BibTeX_file}



\end{document}